 \documentclass[preprint2]{aastex}

\def\cm2{cm$^2$ }
\def\se1{s$^{-1}$ }

\def\arcmin{\hbox{$^\prime$} }
\def\arcsec{\hbox{$^{\prime\prime}$} }

\def\aj{AJ}%
\def\apj{ApJ}%
\def\apjl{ApJ}%
\def\apjs{ApJS}%
\def\ao{Appl.~Opt.}%
\def\aap{A\&A}%
\def\mnras{MNRAS}%
\begin{document}

\title{Kinematics and Modeling of the Inner Region of M\,83}
%\title{Dynamical evolution of the M\,83 central region}
%\title{Strong dynamical evolution in the central region of the bulge of M\,83}

\author{Irapuan Rodrigues\altaffilmark{1,2}, Horacio Dottori\altaffilmark{2,3}, Rub\'en J. D\'{\i}az\altaffilmark{4,5}, Mar\'{\i}a P. Ag\"uero\altaffilmark{5,6} and Dami\'an Mast\altaffilmark{5,6}}

\altaffiltext{1}{IP\&D-Universidade do Vale do Para\'iba -- Av. Shishima Hifumi, 2911 -- 12244-000 - S\~ao Jos\'e dos Campos - SP -- Brazil, irapuan@univap.br}
\altaffiltext{2}{Research Grant from CNPq, Brazil}
\altaffiltext{3}{Dpt. of Astronomy, IF-UFRGS, CP 15051, CEP 91501-970, Porto Alegre, Brazil, dottori@if.ufrgs.br} 
\altaffiltext{4}{Gemini Observatory, Southern Operations Center, La Serena, Chile, rdiaz@gemini.edu}
\altaffiltext{5}{Consejo Nacional de Investigaciones Cient\'{\i}ficas y T\'ecnicas, Argentina}
\altaffiltext{6}{Observatorio Astron\'omico, Universidad Nacional de C\'ordoba, Argentina}

\begin{abstract}
Two-dimensional kinematics of the central region of \objectname{M\,83} (NGC\,5236) were
obtained through three-dimensional NIR spectroscopy with Gemini South telescope. The spatial
region covered by the integral field unit ($\sim 5\arcsec\times13\arcsec$ or
$\sim90\times240\,$pc), was centered approximately at the center of the bulge
isophotes and oriented SE-NW. The Pa${\beta}$ emission at half arcsecond
resolution clearly reveals spider-like diagrams around three centers, indicating
the presence of extended masses, which we describe in terms of Satoh
distributions. One of the mass concentrations is identified as the optical
nucleus (ON), another as the center of the bulge isophotes, similar to the CO
kinematical center (KC), and the third as a condensation hidden at
optical wavelengths (HN), coincident with the largest lobe in 10\,\micron\
emission. We run numerical simulations that take into account ON, KC and HN and
four more clusters, representing the star forming arc at the SW of the optical
nucleus. We show that ON, KC and HN suffer strong evaporation and merge in 10-50
Myr. The star-forming arc is scattered in less than one orbital period, also
falling into the center. Simulations also show that tidal-striping boosts the external shell of the
condensations to their escape velocity. This fact might lead to an
overestimation of the mass of the condensations in kinematical observations with
spatial resolution smaller than the condensations' apparent sizes. Additionally
the existence of two ILR resonances embracing the chain of HII regions, claimed
by different authors, might not exist due to the similarity of the masses of the
different components and the fast dynamical evolution of M\,83  central
300\,pc.

\end{abstract}


%\keywords{Galaxies: nuclei kinematics, stellar composition --- galaxies: individual (NGC 5236)}
%   Modifico para adecuar � revista:
\keywords{Galaxies: nuclei -- galaxies: kinematics and dynamics -- galaxies: stellar content -- galaxies: individual (NGC 5236)}


%% differs from whdiverseat is to be printed in the paper.  The text appearing

\section{Introduction}

\begin{figure*}[htb]
  \begin{center}
    \includegraphics[width=0.9\textwidth,angle=0]{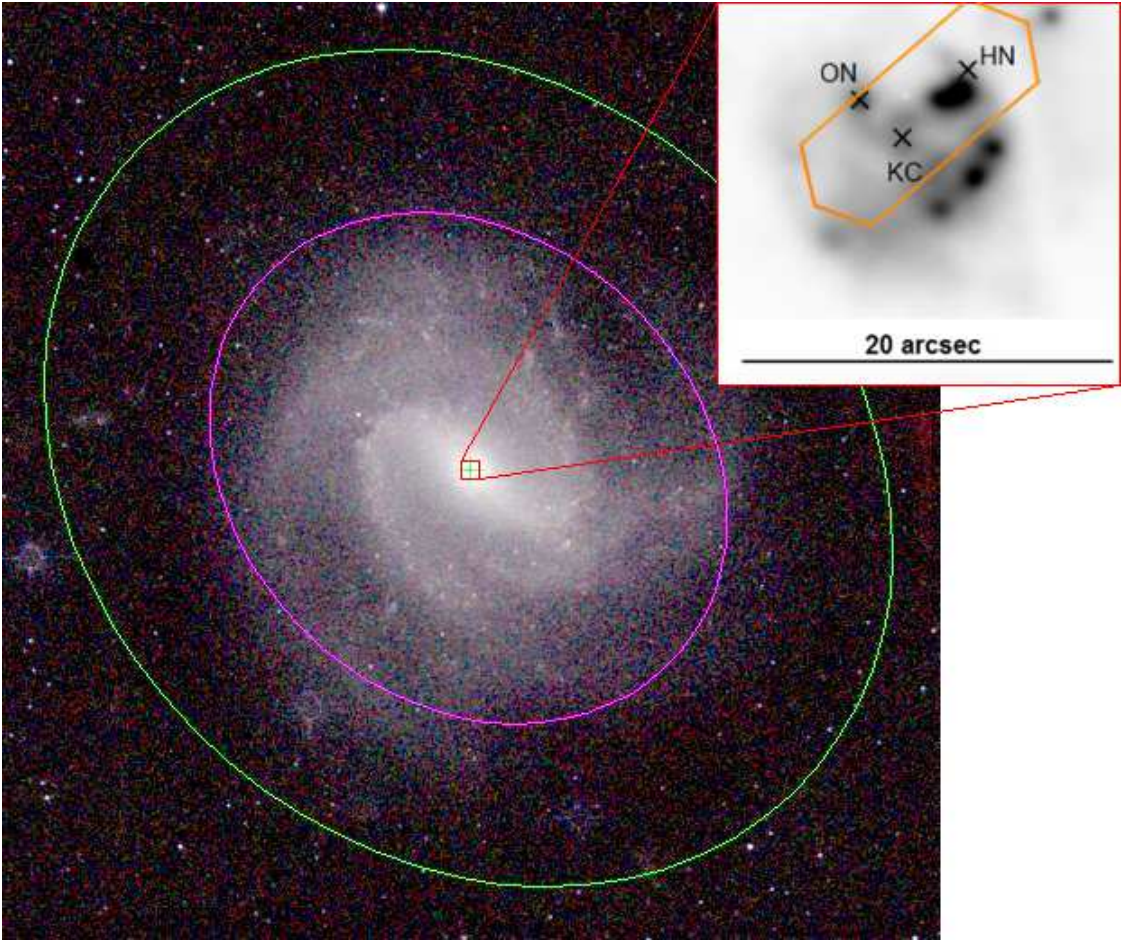}
  \end{center}
  \caption{2-MASS JHK composite image showing two isophotes and a remarkable coincidence
of their center with the bulge center KC. The blowup shows a GMOS/GEMINI-S co-added image of 
the central 20$\times$20 arcsec with the position of KC, ON and HN. North is up and East is to the left. 
CIRPASS field true scale and orientation are outlined in orange}\label{fig:blowup}
\end{figure*}

In spite of the peculiar structure of its circumnuclear region,
\objectname{NGC\,5236} presents a well-defined bulge that can be fitted by de
Vaucouleurs' law in an annular region between radii $\approx 10$\,\arcsec and
40\,\arcsec \citep{1981ApJ...243..716J, 1991A&A...243..309G} corresponding to
$\approx 180 \times 720$ pc. The dramatic inward behavior begins approximately
at a radius of 180-190\,pc, where the main dust lanes associated with the galaxy
bar spiral into a couple of (J-K) rings that, according to
\citet{1998AJ....116.2834E}, might coincide with two inner Linblad resonances.
These rings embrace an arc of about twenty star-forming condensations
\citep{1991A&A...243..309G, 1993BAAS...25..840H, 2001AJ....122.3046H} comparable
to 30\,Doradus in the Large Magellanic Cloud \citep{1993BAAS...25..840H},
clustered in four main knots \citep[Figure 4]{1998AJ....116.2834E}. The center
of this arc is approximately $\approx$ 6.7\arcsec to the SW of the condensation
known as the {\it optical} or {\it visible nucleus} ({\bf ON}) of NGC\,5236. ON
marks the center of the \citet{1998AJ....116.2834E} inner nuclear ring, but not
that of the outer nuclear ring, which is shifted 2.5\arcsec to the SW of the
optical nucleus. K-band spectroscopy \citep{2000A&A...364L..47T} suggests a
dynamical center coincident with the center of symmetry of the K-band external
isophotes ({\bf KC}), which is shifted 4\arcsec to the SW of the optical center
\citep[see Figure\,3 in][and our Figure~\ref{fig:blowup}]{2006AJ....131.1394M}.
This dynamical center is also suggested, although with smaller spatial
resolution, by 2-D CO spectroscopy \citep{2004ApJ...616L..59S}. The region of
Thatte's (2000) isophotal study coincides with that analyzed by
\citet{1981ApJ...243..716J}, and is dominated by the bulge light.

The disk-like appearance of the CO kinematics interior to 300\,pc
\citep{2004ApJ...616L..59S} indicates that the  bar perturbed disk survives well
inside the galaxy bulge. How deep it extends into the nucleus is a question
discussed in this paper. The presence of the hidden condensation ({\bf HN})
\citep{2006ApJ...652.1122D}, which is larger and more massive than the optical
nucleus, raises doubts about the true nature and the role of the optical
nucleus. Furthermore, we also cast doubts on the existence of a double ILR
resonance embracing the arc of HII regions. 
\begin{figure*}[hbt]
  \begin{center}
    \includegraphics[width=0.9\textwidth,angle=0]{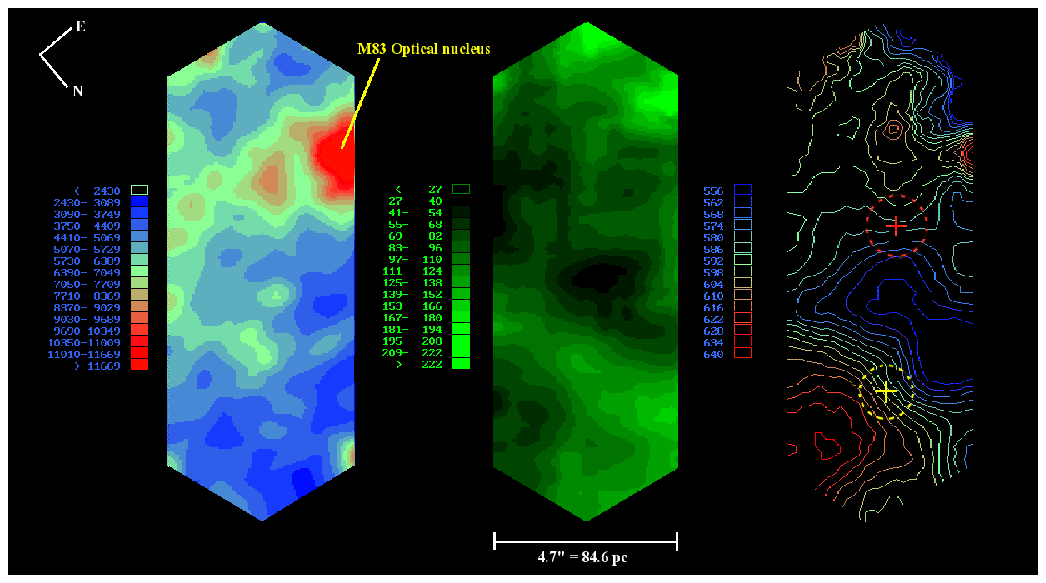}
  \end{center}
  \caption{Panels represent, from left to right, the brightness
  distribution at the Pa${\beta}$ continuum and the Pa${\beta}$ velocity dispersion
and isovelocity maps. The Pa${\beta}$ continuum map shows the position of ON. 
The isovelocity map shows with crosses the positions of KC in red and HN in yellow.
The dashed circles around KC and HN represent the astrometric uncertainty in their position. }\label{fig:velmap}
\end{figure*}

To better understand the richness of phenomena in the central region of
\objectname{NGC\,5236} we performed 3D near infrared spectroscopy at subarsecond
resolution using GEMINI+CIRPASS configuration \citep{2006ApJ...652.1122D}. We
observed a region including the ON, KC and HN. Our study is complemented by
N-body simulations to test the stability of the M\,83 central region, including
the star-forming arc. In Section~\ref{sec:obs} we present our observations, in
Section~\ref{sec:kin} we discuss the central kinematics and the three most
important condensations, which present differentiated kinematics. Numerical
simulations are presented in Section~\ref{sec:simul} and in
Section~\ref{sec:conclu} we present our conclusions.

\begin{table*}[hbt]
  \begin{center}
  \caption{Condensation data. The $2\sigma$ uncertainties in the astrometry of ON, KC
and HN are $0.15\arcsec$, $0.8\arcsec$ and $0.7\arcsec$, respectively. The three last columns 
show characteristics of the disk around each one of the condensations: R$_{max}$
is the largest extension along the line of the nodes where disk kinematics can be reliably distinguished. V$_{peak}$ is the peak velocity along the line of the nodes, corrected for inclination and
$\sigma$ is the velocity dispersion at the position of the condensations.}
  \label{tab:kindat}
  \begin{tabular}{cccccc}
    \tableline\tableline
                & RA       & Dec.      & R$_{max}$   & V$_{peak}$  & $\sigma$  
\\
                &  (J2000) &   (J2000)  & [pc]        & [km\,s$^{-1}$]     & [km\,s$^{-1}$]    
\\
    \tableline
       KC       & $13^h37^m0.56^s$ & $-29^{\circ}\ 51\arcmin\ 56.9\arcsec $ &
$65\pm5$ &$68\pm8$  & $82\pm10$  \\
       ON       & $13^h37^m0.98^s$ & $-29^{\circ}\ 51\arcmin\ 55.5\arcsec $ &
$8\pm1$  & $46\pm9$  & $110\pm10$ \\
       HN       & $13^h37^m0.46^s$ & $-29^{\circ}\ 51\arcmin\ 53.6\arcsec $ &
$45\pm8$ &$58\pm11$ & $96\pm10$  \\
    \tableline
  \end{tabular}
 \end{center}
\end{table*}

\section{Observations}\label{sec:obs}

We used the {\it Cambridge IR Panoramic Survey Spectrograph}
\citep{2000SPIE.4008.1193P, 2004SPIE.5492.1135P} installed on the GEMINI South
telescope in March 2003. These observations, performed in queue mode, were taken
with an integral field unit (IFU) sampling of 0.36\arcsec (6.4\,pc) with an
array with size $\approx 5\arcsec \times 13\arcsec$. The array of 490 hexagonal
doublet lenses attached to fibers provides an area filling factor of nearly
100\%. The IFU was oriented at PA\,$=120^{\circ}$\ and centered slightly to the
NW of the kinematical center according to \citet{2002BAAA...45...74M} position.
In Figure~\ref{fig:blowup}, a sketch of the detector position is shown. It was
exposed for 45 minutes and covers a spectral range between $1.2-1.4\,\micron$,
including Pa$\beta$\,1.3\,\micron\ and [FeII]\,1.26\,\micron, with a spectral
resolution of approximately 3200. The seeing was about 0.5$^{\prime\prime}$;
therefore, the focal plane was slightly sub-sampled by the configuration
constraint.

The data were reduced using IRAF (distributed by the NOAO), ADHOC (2D kinematics
analysis software developed by Marseille's Observatory) and SAO (spectra
processing software developed by the Special Astrophysical Observatory, Russia).
More details on the reduction are presented in \citet{2006NewAR..49..547D} and
the general technique is thoroughly discussed by \citet{1999ApJ...512..623D} and
\citet{2002BAAA...45...74M}.

\section{2-D kinematics of M\,83 central region at less than 10\,pc spatial resolution}\label{sec:kin}

In Figure~\ref{fig:blowup}, we present the composed JHK 2-mass image of M\,83.
The inset shows a blow-up of the central 400$\times$400\,pc, where we detach the
positions of ON, KC and HN. The astrometric positions of the condensations are
given in  Table~\ref{tab:kindat} \citep[the astrometry is discussed
in][]{2006NewAR..49..547D}. In Figure~\ref{fig:velmap}, we show the brightness
distribution at the Pa$\beta$ continuum, the Pa$\beta$ radial isovelocity
contours, and the Pa$\beta$ velocity dispersion map. The velocity dispersion is
shown  as the line FWHM (Table~\ref{tab:disks}). In spite of the complex
kinematics shown by the 2-D radial velocity map, we can distinguish ordered
radial velocities in the form of spider diagrams around ON, KC and HN. As is
well known, a spider diagram indicates disk-like motions. The sizes of the disks
around the three condensations, their maximum radial velocities along the line
of the nodes, and their velocity dispersions at their centers of symmetry are
presented in Table~\ref{tab:kindat}.

\subsection{The ionized gas kinematics around KC}\label{sec:kc}
% R$\,=\,3.5\arcsec\ \pm\ 0.8\arcsec $, P.A.$ = 249^{\circ}\ \pm 3^{\circ}$

\begin{figure}[bht]
  \begin{center}
    \includegraphics[width=0.45\textwidth,angle=0]{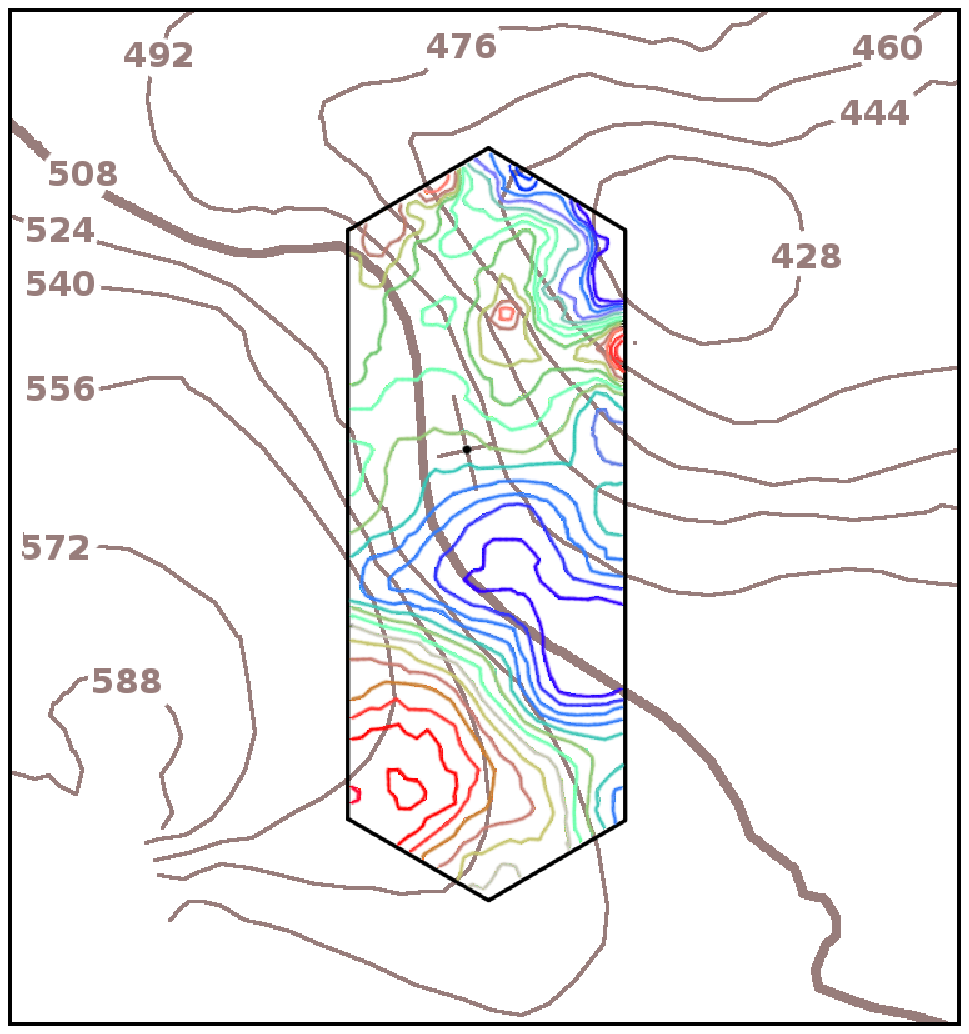}
  \end{center}
  \caption{Central part of the~\citet{2004ApJ...616L..59S} CO (2--1) velocity map (brown contours) superimposed on the Pa${\beta}$ isovelocity maps. }\label{fig:zerovel}
\end{figure}

KC is the most extended of the three disks. Its center coincides, up to the
margin of error, with the center of symmetry of the bulge isophotes
\citep{1981ApJ...243..716J,2000A&A...364L..47T}, which in turn coincides with
that of the CO kinematical center \citep{2004ApJ...616L..59S}. We recall that
the GEMINI spatial resolution is at least twenty times higher than that of
Sakamoto's CO observation.

The major axis of the spider diagram is at P.A.$=120^{\circ}$. From the spider
diagram major-to-minor axis ratio, we derive an inclination of $68^{\circ} \pm 8
^{\circ}$ for the gaseous disk. We have to be cautious at this point because the
minor axis velocity field seems to be perturbed. The same situation was faced by
\citet{2004ApJ...616L..59S}, who also quote two highly discrepant possible
values for the  inclination of  the 300\,pc CO disk. These authors finally chose
 the inclination angle of the large scale galactic disk based on kinematical
reasons. In our case, the best two-dimensional velocity model fitting to the
observed velocity field is attained for an inclination angle of $25^{\circ} \pm
8^{\circ}$ (see Section~\ref{sec:obs-mod}), which coincides, up to the margin of
error, with the small inclination of the large scale M83 disk. A small angle
also better fits the innermost scale rotation curves to Sakamoto's 300\,pc disk
one (see Fig.3). The inflow of matter along the bar into the central region
might be mimicking the disks' high inclination. The Pa$\beta$ disk around the KC
and Sakamoto's  CO disk present different orientations (see
Figure~\ref{fig:zerovel}), probably pointing to a phenomenon of transition
between $x_2$- and $x_1$-like orbits, as quoted by \citet{2004ApJ...616L..59S},
or to a strong perturbation of the central structure, as discussed by
\citet{2006ApJ...652.1122D}. 

The spider diagram (see Figure~\ref{fig:velmap}) can be traced up to R$_{max}\approx$\,65\,pc but it
is strongly perturbed to the E at the receding extreme, and to the W at the
approaching extreme. The KC rotation curve was obtained from approximately 60 radial
velocity points distributed along the whole disk. From this dataset we obtained
13 mean rotation velocities between 4\,pc and 65\,pc. Each velocity point
entering the mean was weighted by $w(\alpha)=\,|cos(\alpha)|$, where $\alpha$ is
the position angle of the line joining the point to the disk center with respect
to the line of the nodes. The disk around KC can be described in terms of a
Satoh's-like spheroid  with an effective radius of R$_{KC}=38 \pm 8$\,pc and a
mass inside R$_{max}$ of M$_{KC}\approx (60.0\pm 20) \times 10^6$M$_\odot$ 
(for a disk inclination of $25^{\circ}$). We note that an inclination of $68^{\circ}$ 
for the disk around KC would give a mass of $\approx (17\pm 2) \times 10^6$M$_\odot$, 
similar to that deduced spectroscopically by \citet{2000A&A...364L..47T} 
from the $^{12}$CO bandhead velocity broadening.

The velocity dispersion in the non-resolved central part is of the order of
80\,km\,s$^{-1}$ (see middle panel in Figure~\ref{fig:velmap}). Assuming that this
velocity dispersion is due to a central mass concentration, namely a putative
BH, we determined the BH mass upper limit, M$_{BH}$, by adding
Keplerian rotation curves convolved with a 9 pc Gaussian to that of the Satoh
disk, and constraining the resulting Satoh+Kepler rotation curve with the central
velocity dispersion and the errors of the measured velocity points. The result
of this procedure is shown in Figure~\ref{fig:bhrot} for M$_{BH}=0.2, 1.0$ and
$2.0\times10^6$\,M$_{\odot}$. The best fit is quoted in Table~\ref{tab:disks}.

This analysis  points to KC as the true nucleus of M\,83.

\begin{table}[htb]
  \begin{center}
  \begin{small}
  \caption{Satoh model-fitting parameters.}
  \label{tab:disks}
  \begin{tabular}{llll}
    \tableline\tableline
			      & KC         & ON        & HN	        \\
    \tableline
PA         $[^{\circ}]$	      & 120	     & 120       & 64           \\
i          $[^{\circ}]$	      & $25.0\pm8$   & $60\pm10$ & $62\pm8$     \\
R$_{eff}$  [pc]	      	      & $38.0\pm4$   & $8\pm1$   & $33\pm3$     \\
$M_K$      [$10^6 M_{\odot}$] & $60.0\pm20$  & $4\pm2$   & $20\pm7$ 	\\
$M_{BH}$   [$10^6 M_{\odot}$] & 0.2 -- 1.0   & $\le 1.0$ & 0.2 -- 1.0   \\
    \tableline
  \end{tabular}
 \end{small}
 \end{center}
\end{table}

\begin{figure*}[htb]
\begin{center}
\includegraphics[width=0.48\textwidth]{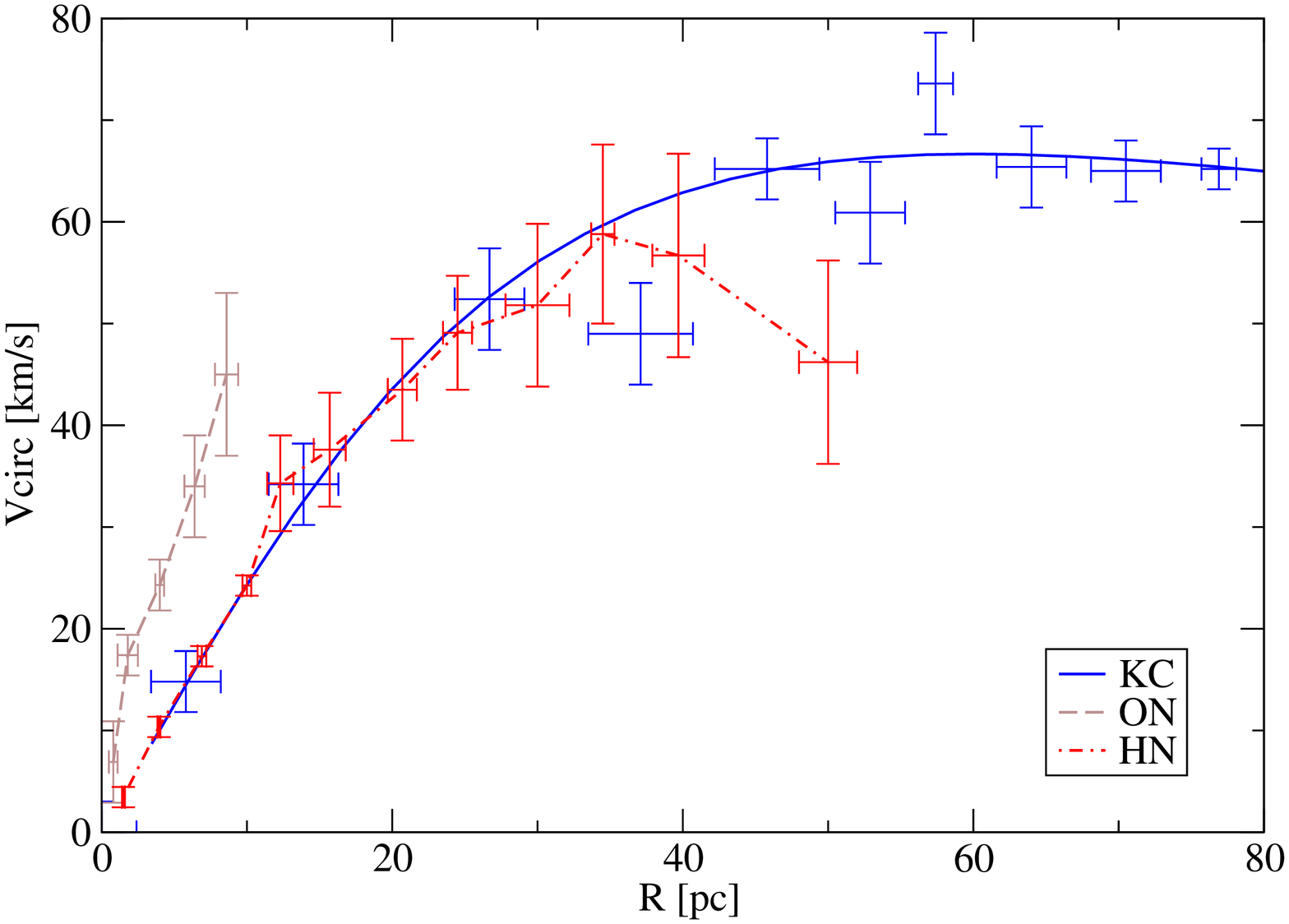}
\includegraphics[width=0.48\textwidth]{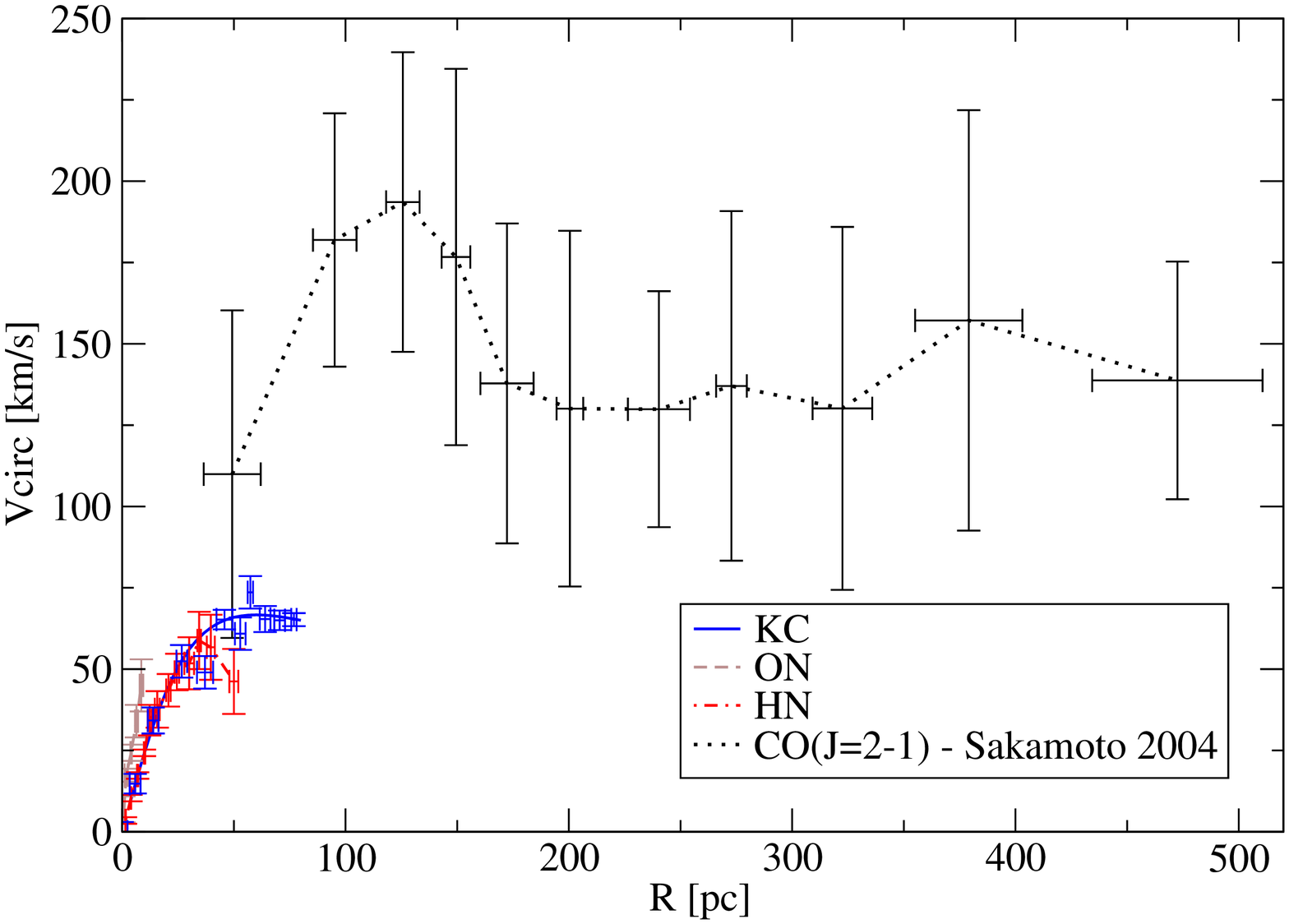}
\end{center}
  \caption{Rotation curves of the Kinematical Center (KC), the Optical
  Nucleus (ON) and the Hidden nucleus (HN). They are compared in the right panel to the 300\,pc 
CO disk rotation curve, as obtained from \citet{2004ApJ...616L..59S}. } \label{fig:rot}
\end{figure*}

\begin{figure*}[hbt]
\begin{center}
\includegraphics[width=0.48\textwidth]{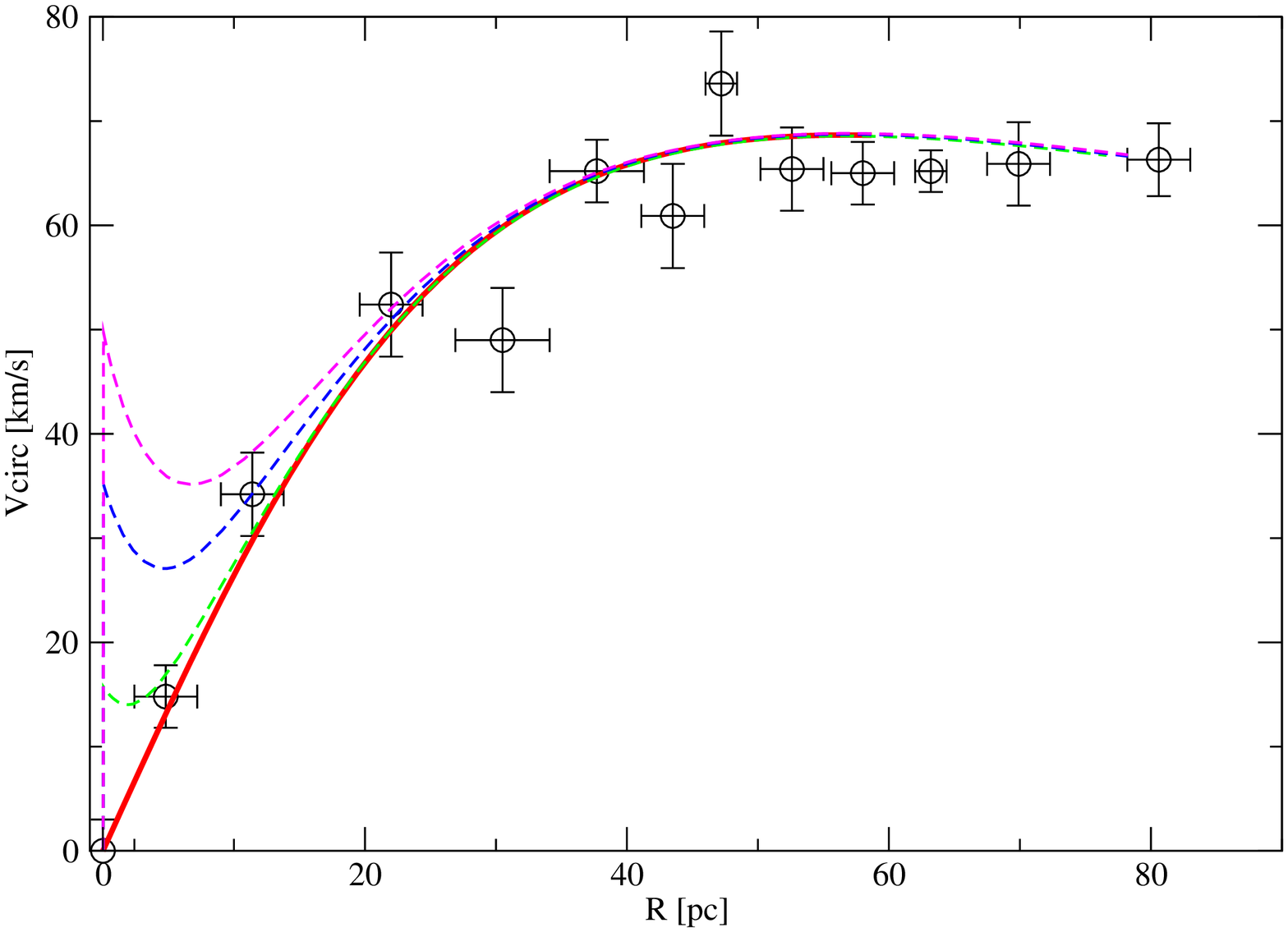}
\includegraphics[width=0.48\textwidth]{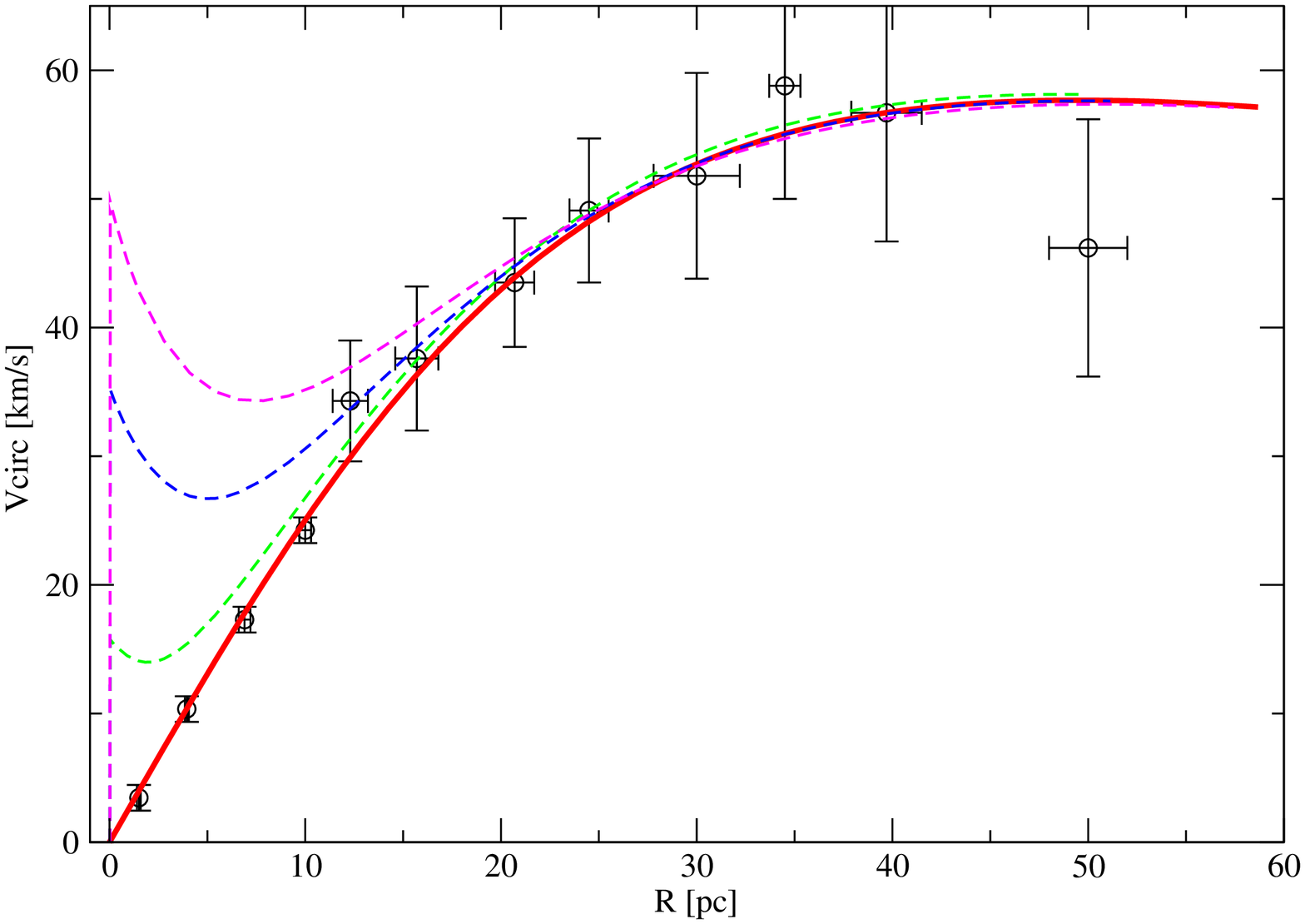}
\end{center}
  \caption{The rotation curves of (a) KC and (b) HN were fitted
  by Satoh's models (red line). The dashed lines (green, blue and magenta in the electronic edition) were
  obtained by adding point-like sources of M$_{BH}=0.2, 1.0$ and
  $2.0\times10^6$\,M$_{\odot}$, convolved with a 9 pc
  Gaussian. M$_{BH}=0.2\times10^6$\,M$_{\odot}$ is more compatible in
  both cases.}\label{fig:bhrot}
\end{figure*}

\subsection{The ionized gas kinematics around HN}\label{sev:hn}

Located at R\,$=7.8\arcsec \pm 0.7\arcsec$ to the NW of KC (Fig. 2),  the spider
diagram of the ionized gas around HN  presents a P.A.\,$=64^{\circ}$. The form
of the spider diagram (open legs) points to a more concentrated and more
homogeneous distribution than that of the KC. It is probably a cannibalized
dwarf satellite of M\,83, as discussed by \citet{2006NewAR..49..547D}. To derive
the rotation curve as for KC, we used a distribution of more than 50 points
along the whole disk, which provided 13  mean rotation velocities up to
R\,$=50$\,pc. In Tables \ref{tab:kindat} and \ref{tab:disks} we present the
kinematical data for HN and the derived Satoh disk parameters. The mass of HN
was also derived from the rotation curve, as well as that of the putative Black
Hole at its center (see Table~\ref{tab:disks}), through the velocity dispersion,
following  the procedure previously outlined for KC.

\subsection{The ionized gas kinematics around ON}\label{sec:on}

ON is traditionally referred to as the off-centered optical nucleus
of M83. \citet{2004ApJ...616L..59S} proposed that it could be
a cannibalized satellite. 

From photometric data, \citet{1998AJ....116.2834E} derived a mass of $4 \times
10^{6}$ M$_\odot$ and \citet{2000A&A...364L..47T} of $2.5 \times 10^{6}$
M$_\odot$. From $^{12}$CO band-head 2.293\,$\mu m$ velocity dispersion,
\citet{2000A&A...364L..47T} derived a kinematical mass of $1.3 \times 10^7$
M$_\odot$ within 5.4 pc, assuming that the optical nucleus is virialized. The
mass of ON derived photometrically disagrees by a factor of 4 to 5 from that
derived spectroscopically, probably indicating a system with a non-virialized
periphery, as discussed in Section~\ref{sec:simul} in connection with the
many-body interaction. The disk around ON is the smallest of the three disks
discussed in this paper. The ON kinematical center is slightly shifted to the SW
of the ON continuum peak at $1.3\,\mu m$, as can be seen in
Figure~\ref{fig:velmap}. The velocity gradient across ON is determined from data
near the spatial resolution limit. We are probably dealing with problems of
oversampling. The rotation curve with mean errors is shown in
Figure~\ref{fig:rot}. The mass of ON was also derived from the rotation curve,
as well as that of the putative Black Hole at its center (see
Table~\ref{tab:disks}), using the velocity dispersion and the procedure
previously outlined for KC. Our determination of the mass agrees very well with
that obtained from photometric data.

\subsection{Fitting the whole central radial velocity field of the gas}\label{sec:obs-mod}

\begin{figure}[thb]
	\begin{center}
		(a)~~~~~~~~~~~~~~~~~~~~~~~~~~~~~~(b) \\
		\includegraphics[width=0.23\textwidth]{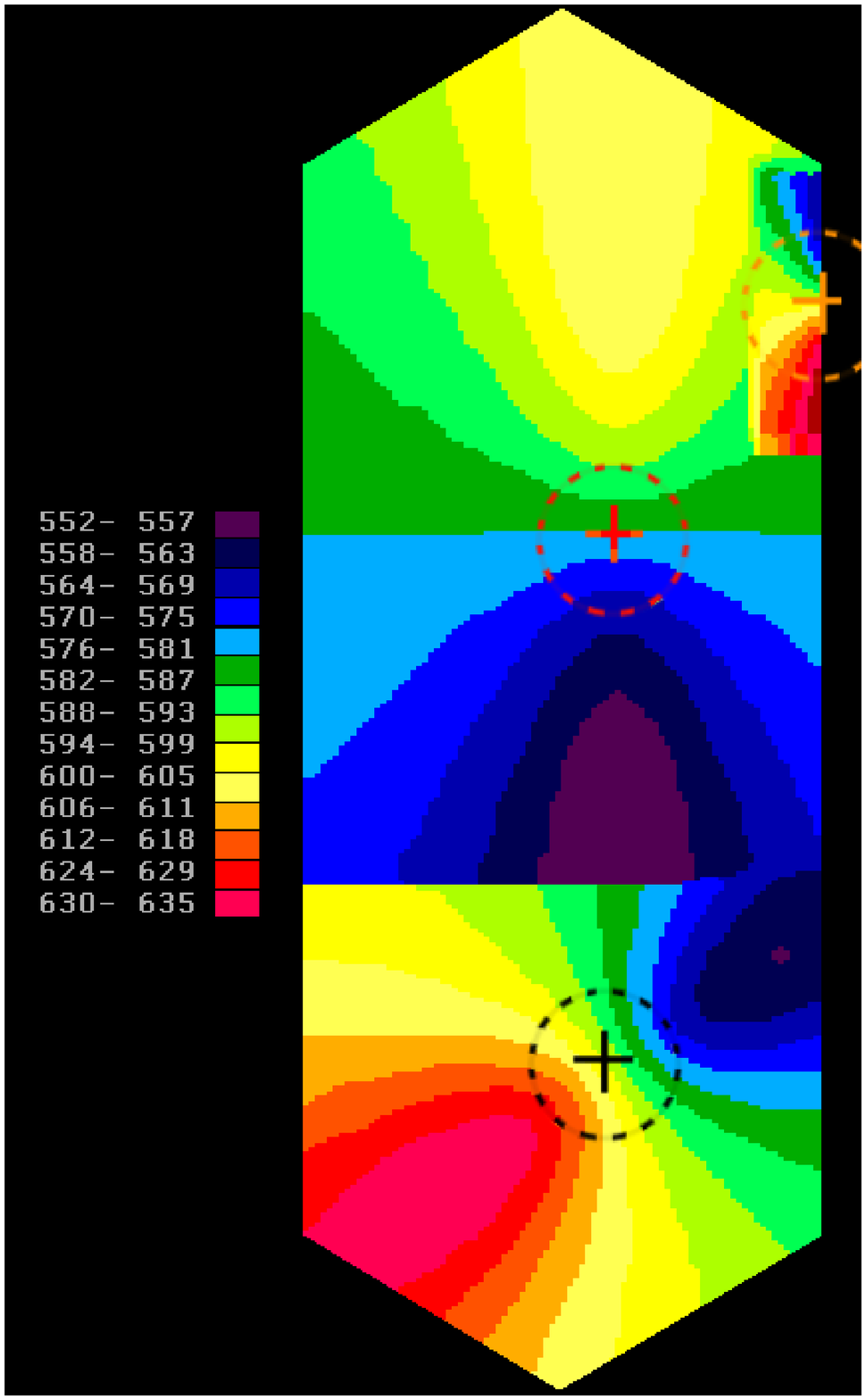}
		\includegraphics[width=0.2246\textwidth]{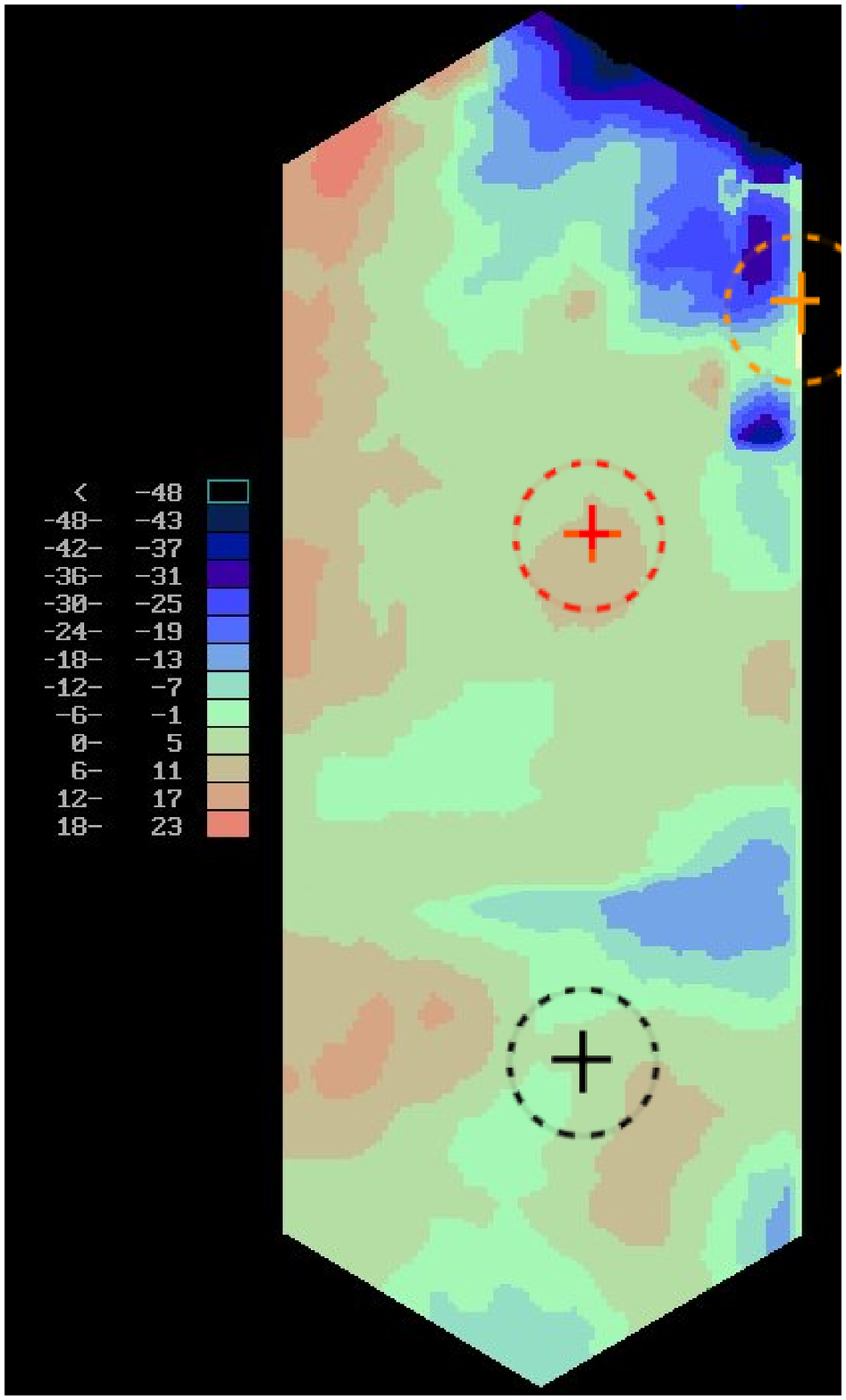}
	\end{center}
	\caption{(a) Velocity map model obtained by superposing the disks around KC, HN and KC. (b) Residual velocity map resulting from the subtraction of the modeled velocity map from the observed one.}\label{fig:obs-mod}
\end{figure}

The data in Tables~\ref{tab:kindat} and~\ref{tab:disks} allow us to generate a
composed radial velocity map model (Figure~\ref{fig:obs-mod}a), to be compared
with the observed one. The fitting of KC was the most critical due to the
difficulty in the determination  of its disk inclination angle (\ref{sec:kc}).
Several fits were made, keeping constant the inclination angles of the disks
around ON and HN and allowing the one associated with KC to vary from $25^{\circ}$ to
$60^{\circ}$. We verified that the disk-like residual around KC decreases
dramatically when approaching smaller inclinations. Finally, the smallest
residuals are obtained for an inclination of $25^{\circ}\pm 5^{\circ}$. The
final map is shown in Figure~\ref{fig:obs-mod}b. This inclination is in good
agreement with the inclination of the large scale M83 disk and that of the
inner 300\,pc, derived from CO kinematics \citep{2004ApJ...616L..59S}, as can be
seen in Figure~\ref{fig:rot}. This again points to a continuity of the galactic
disk kinematics from a kiloparsec scale to the smaller scales studied in
this paper, although variations in the position angles of the lines of nodes are
observed at 300\,pc and at scales smaller that 100\,pc as well (see
Figure~\ref{fig:zerovel}). 

Figure~\ref{fig:obs-mod}b shows that around KC, ON and
HN the residuals are mainly at the noise  level, indicating that the interpretation
of the observed velocity map as resulting from the superposition of three disks
is, 
% Mudanças sugeridas pelo referee
from a qualitative point of view, essentially correct. Nevertheless, deviations from rotation points to a more complex situation and indicates that the mass estimations based on rotation curves are barely aproximate. As an example we
% Fim da mudança
detect anomalous blueshifted kinematics at the East
extreme (upper right) above ON, and redshifted kinematics to the West of HN. These regions
coincide with the high velocity dispersion zones in Figure~\ref{fig:velmap}. These
might indicate regions of inflow of gas along the bar falling into the central
region.

% referee's last suggestion:
% We should note that in the spider diagrams presented, there is much more than only
% rotation; there are deviations from rotation indicating the complexity
% of the situation and that the mass estimations based on the rotation
% curve analysis are barely approximate. The model could change if this is considered. 
% Probably this would not change qualitatively the results but it could
% change them quantitatively.

The fit of the rotation curve, from the scale of tens to hundreds of parsecs
(Fig.3), does not seem to confirm the existence of double ILR, as claimed by
\citet{1998AJ....116.2834E} and \citet{2004ApJ...616L..59S}.

\begin{figure*}[bht]
\begin{center}
\includegraphics[width=0.3\textwidth]{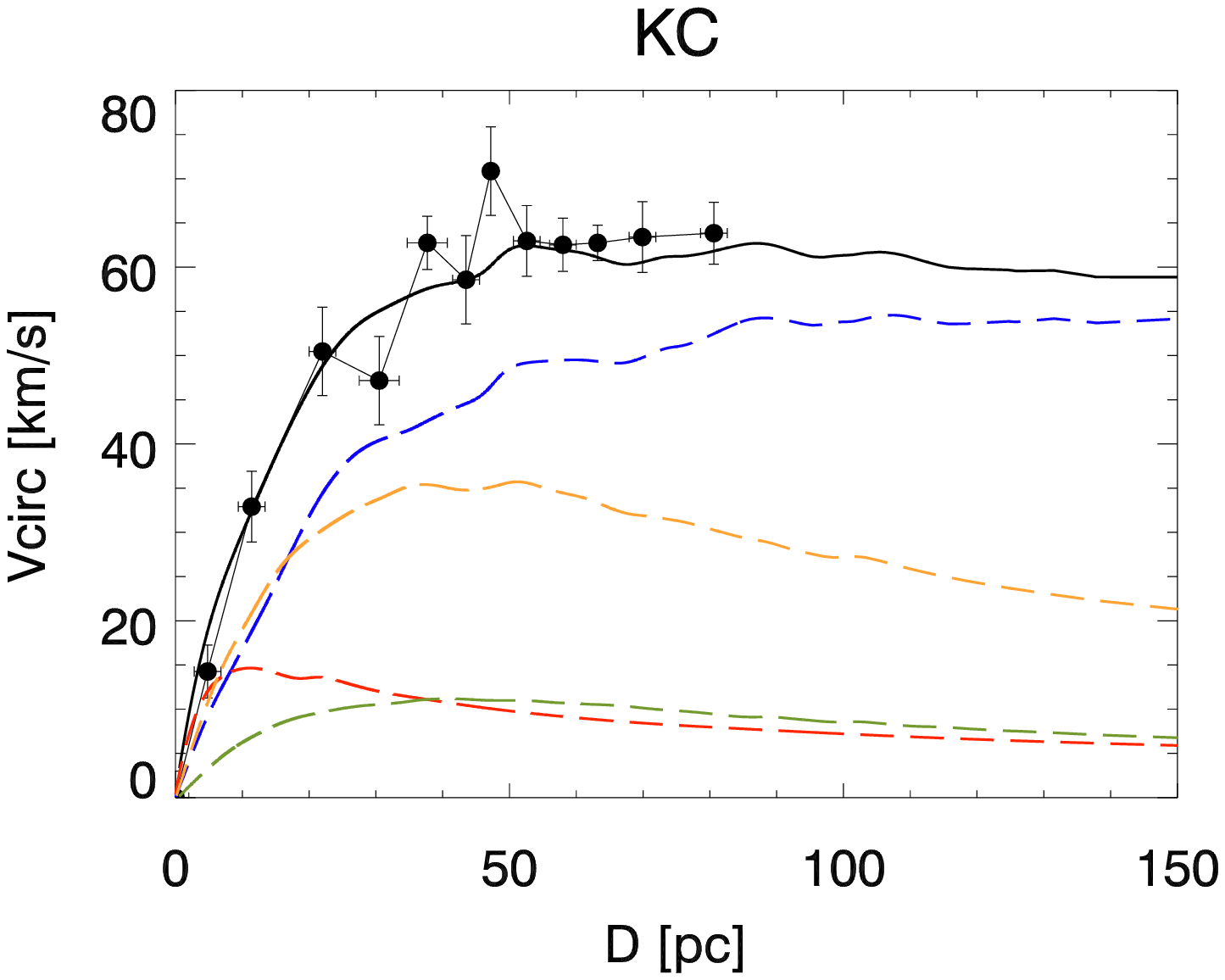}
\includegraphics[width=0.3\textwidth]{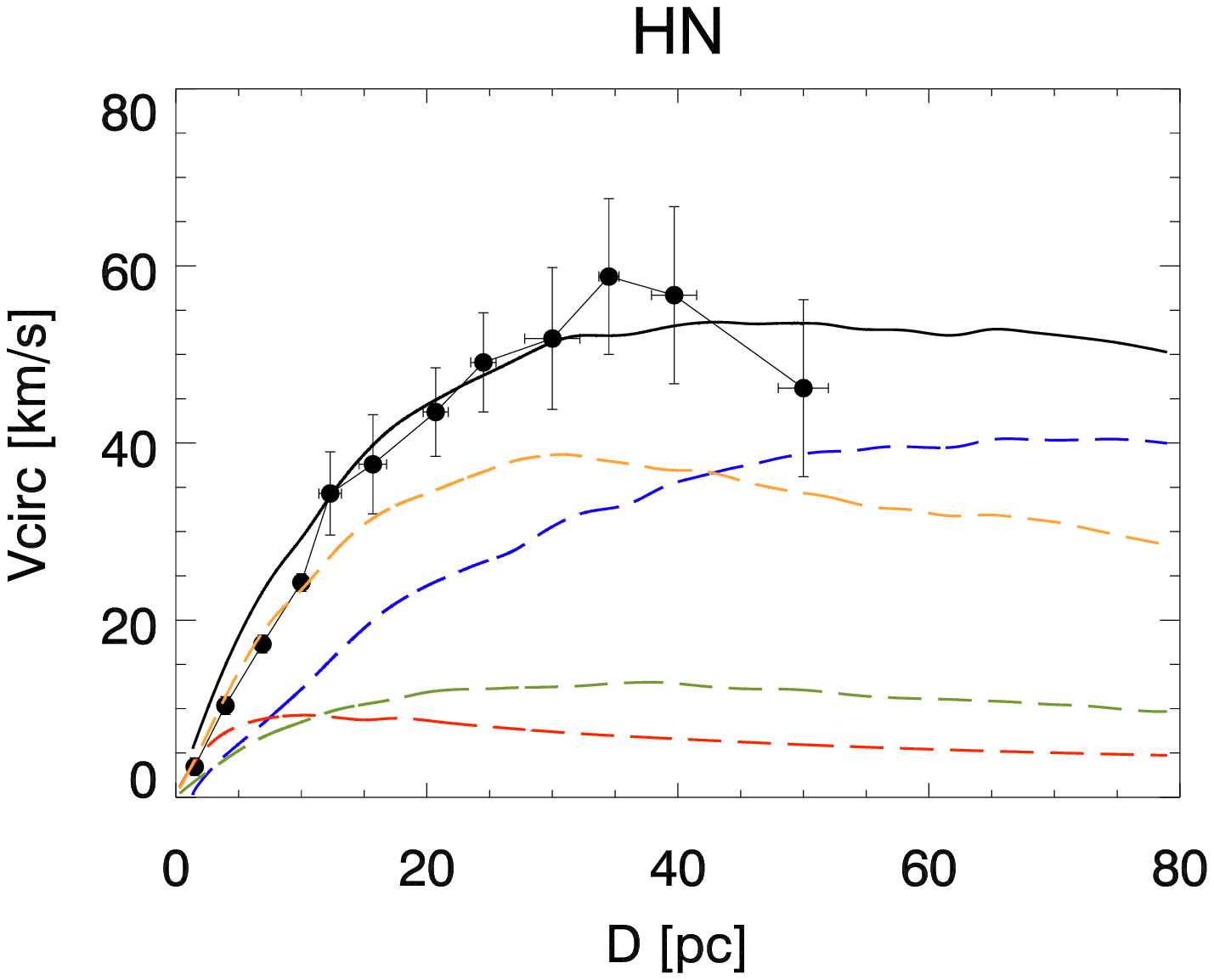}
\includegraphics[width=0.3\textwidth]{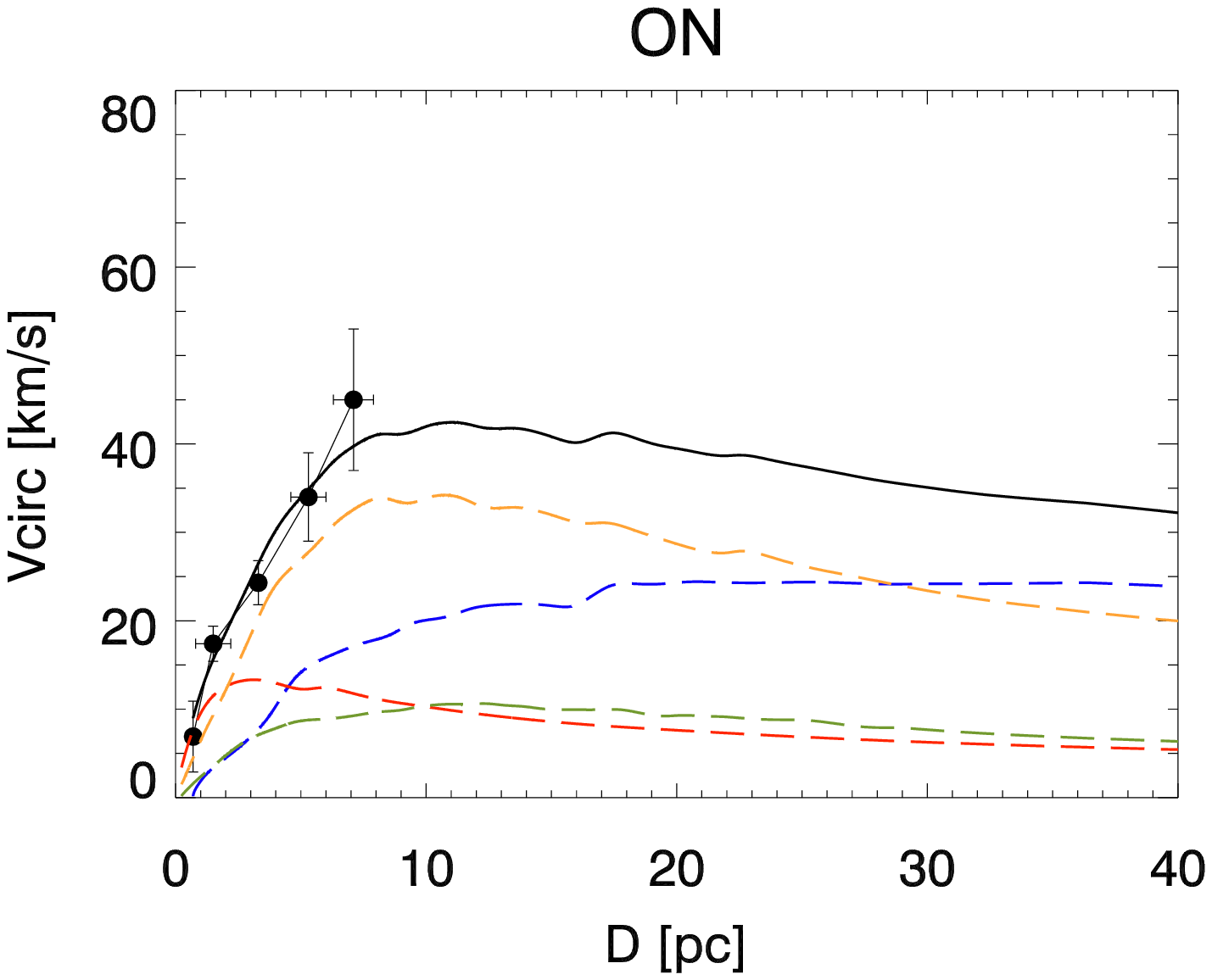}
\end{center}
\caption{From left to right: Rotation curves of KC, HN and ON modeled with Hernquist compound models. Dashed lines are separate components and continuous line is the model circular speed.
Black dots with error bars are from CIRPASS data.}\label{fig:modrotcurve}
\end{figure*}

\section{N-body simulations}\label{sec:simul}

In order to understand the dynamical evolution of the M\,83 central region, we
performed N-body simulations of a system composed of KC, HN, ON and four
condensations representing the star-forming arc. Hernquist's models were used
for KC, HN and ON \citep{1993ApJS...86..389H}. We included stellar and gaseous
components. The models were constrained by the bodies' rotation curves.
Figure~\ref{fig:modrotcurve} shows the model rotation curves adjusted in order
to fit the observational data. The model parameters are quoted in Table
\ref{tab:models}. The four condensations on the star forming arc were
represented by Plummer's models, with 1000 particles each, with positions and
masses corresponding to regions 2 to 5 in Figure\,8 of
\citet{1998AJ....116.2834E}, where:  region 2 has M$_2 = 1.3 \times 10^6$
M$_\odot$; regions 3 and 4 have M$_3 =$ M$_4 = 1.4 \times 10^6$ M$_\odot$; and
region 5 has M$_5 = 1.8 \times 10^6$ M$_\odot$. All Plummer models have a core
radius of 1.5\,pc and a cutoff radius of 15\,pc. 

A large set of simulations was run in order to refine the models and orbital
parameters used in the final simulation. The orbital parameters were chosen
considering that M83 rotates clockwise. Inclinations and position angles are as
in Table\,2. Here we suppose that clusters 2 to 5 on the star forming arc, as
well as ON are in circular clockwise (prograde) orbits in the plane of the
external disk of M83. According to \citep{2006ApJ...652.1122D}, HN is the
remnant of a captured galaxy that triggered the star formation along the arc of
HII regions. The form of the arc and its position with respect to HN lead us to
adopt a circular counter-clockwise orbit for HN.

The simulations were performed with Gadget2 \citep{2005MNRAS.364.1105S}. A total
of 91,552 particles was used, and the simulations were run for 300 Myr, starting
at the present time configuration. We considered starting at a previous epoch, 
before the formation of the arc of HII regions, but the complexity of the
environment at the M83 central region and the number of parameters required to
qualitatively  take into account all possible configurations in the past would be
a matter for a paper especially devoted to that study. 

\begin{figure*}[htb]
\begin{center}
\includegraphics[width=0.32\textwidth]{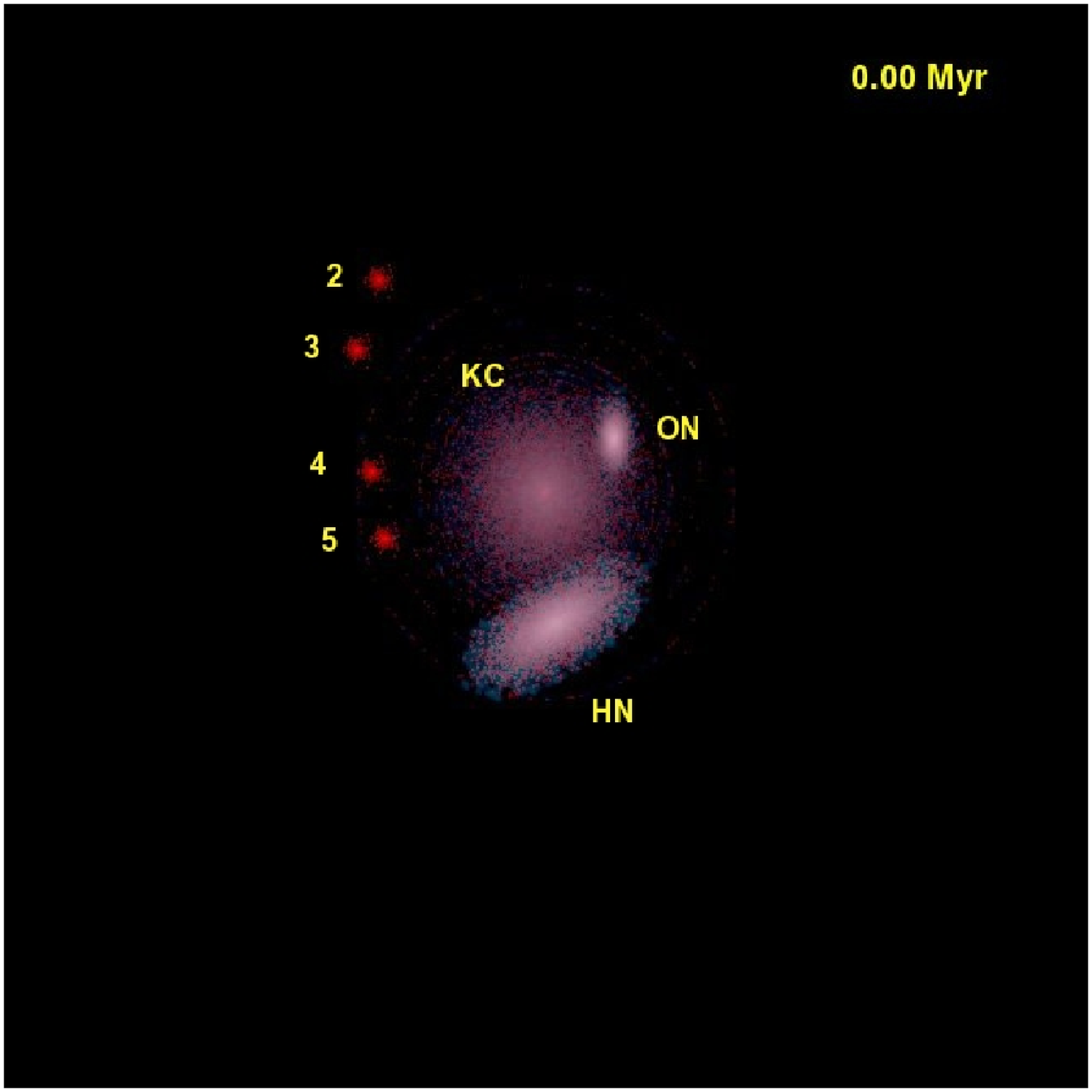}
\includegraphics[width=0.32\textwidth]{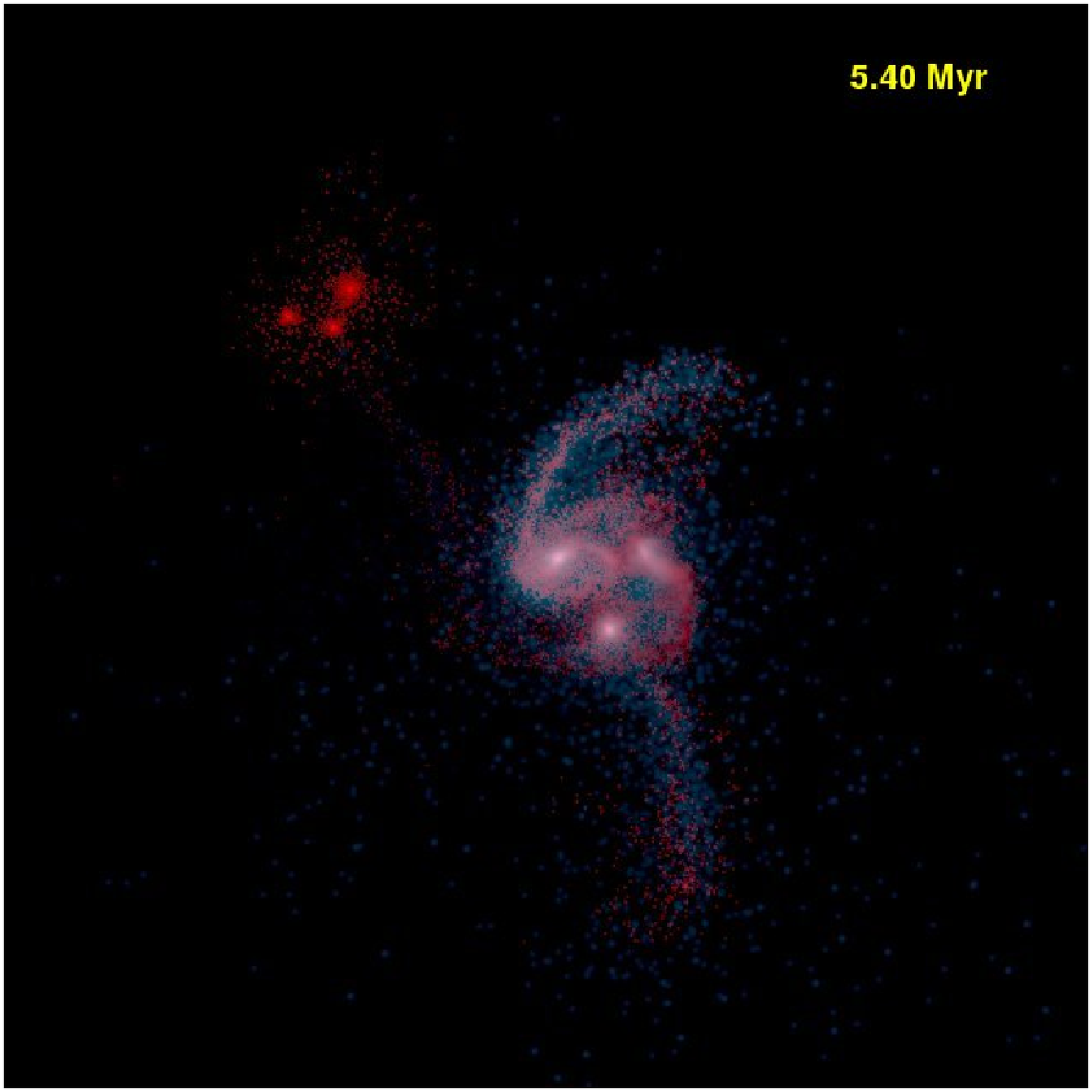}
\includegraphics[width=0.32\textwidth]{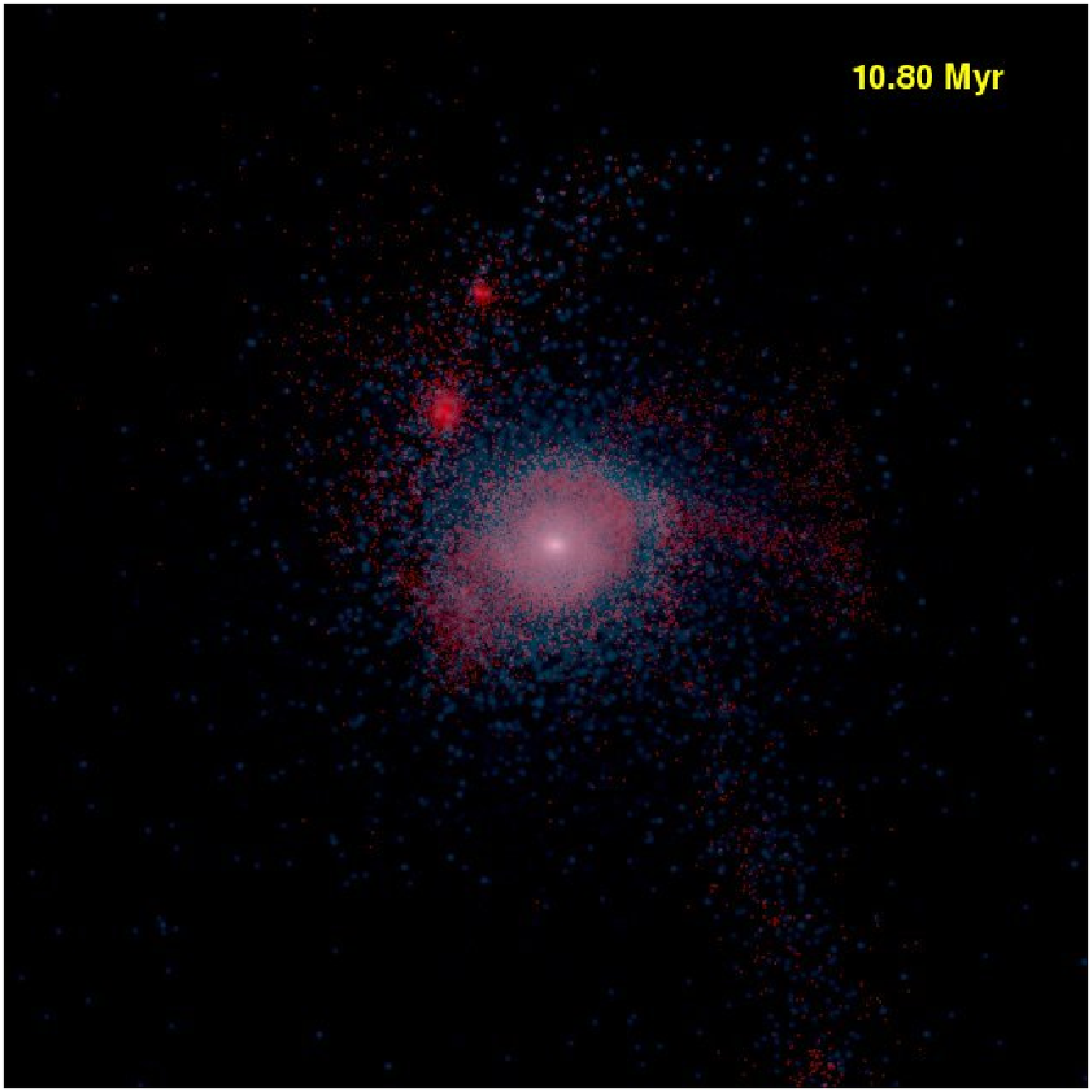}
\includegraphics[width=0.32\textwidth]{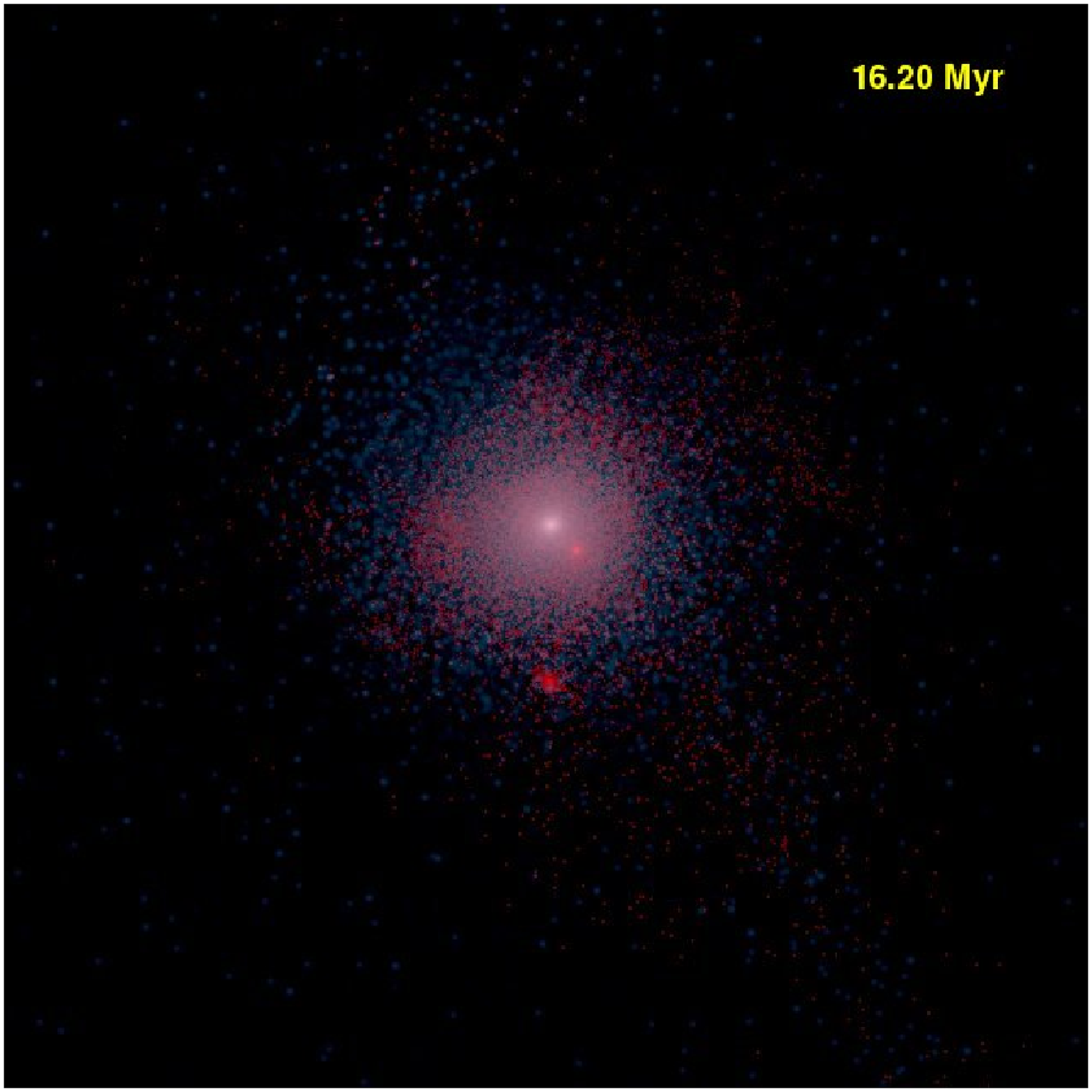}
\includegraphics[width=0.32\textwidth]{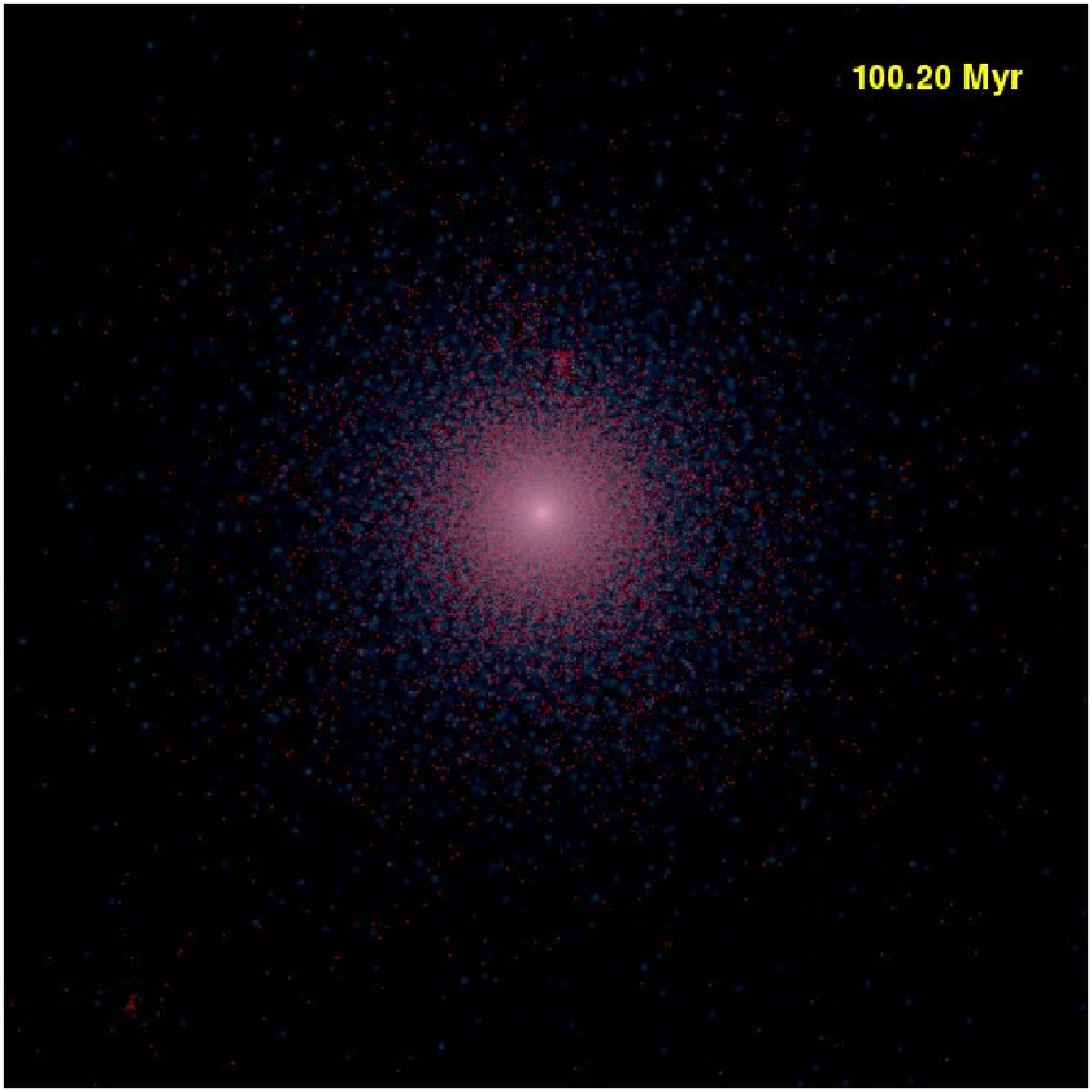}
\includegraphics[width=0.32\textwidth]{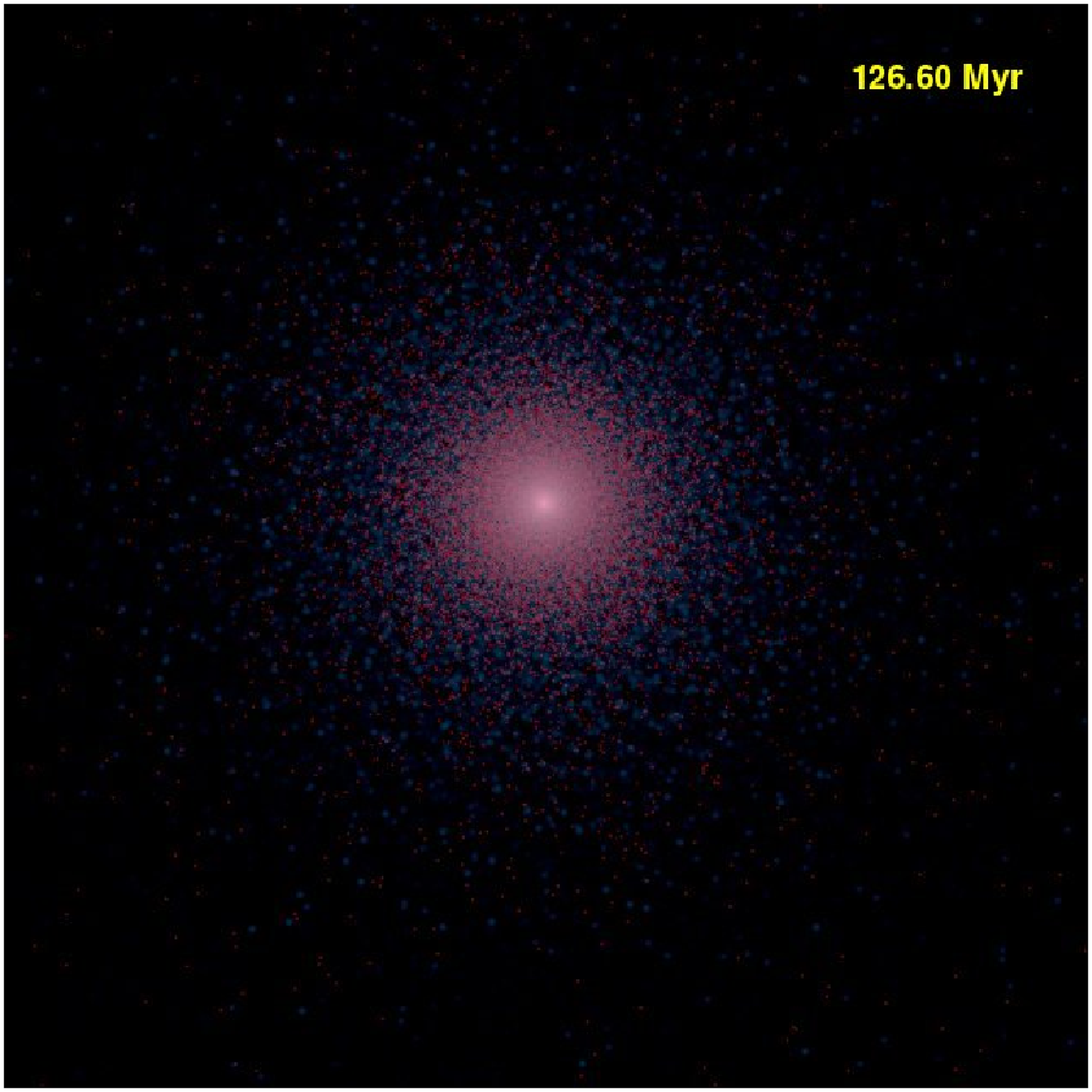}
\end{center}

\caption{Evolution of the M83 central region model. ON and the Plummer models (2
to 5) are in prograde (clockwise) orbits around KC, while HN is in retrograde
orbit. KC, HN and ON merge in 16 Myr. The last Plummer sphere (cluster 4)
coalesces with the central body in 130 Myr. Gas particles are blue, stellar
particles red. Spherical components of KC, HN and ON models are not shown for
clarity.}\label{fig:sim}

\end{figure*}

The simulation (see Figure \ref{fig:sim}) shows that the galaxy
nucleus (KC), the optical nucleus (ON) and the hidden nucleus (HN)
would form a single massive core in about 16 Myr. All 
Plummer models (clusters 2 to 5) fall into the
nucleus in 130 Myr.

\begin{table*}[htb]
 \begin{center}
  \caption{KC, HN and ON model parameters.}  \label{tab:models}
 \begin{small}
  \begin{tabular}{l c c c}
   \tableline\tableline
                                	&{\bf KC model}	&{\bf HN model} &{\bf ON model}
\\
   \tableline
Number of particles in disk        	& 16384		&   8192	&   4096	\\
Disk mass               		& 15.0		&   13.0	&   3.5		\\
Disk radial scale length        	& 17.0		&   13.0 	&   4.5		\\
Disk vertical scale thickness		& 1.7		&   0.8		&   0.4		\\
Reference radius R$_{ref}$      	& 42.5		&   30		&   8.0		\\
Toomre Q at R$_{ref}$           	& 1.5		&   1.5		&   1.5		\\
	\tableline
Number of particles in gas disk		& 16384		&   8192	&   4096	\\
Gas disk mass               		& 1.5		&   1.5		&   0.35	\\
Gas disk radial scale length		& 17.0		&   13.0	&   4.5		\\
Gas disk vertical scale thickness	& 1.7		&   0.8		&   0.4		\\
Toomre Q at R$_{ref}$           	& 1.5		&   1.5		&   1.5		\\
\tableline
Number of particles in bulge       	& 512		&   512		&   512		\\
Bulge mass              		& 1.3		&   0.5		&   0.3	\\
Bulge radial scale length       	& 3.4		&   4.0		&   1.0		\\
\tableline
Number of particles in spherical component  	& 16384		&   8192	&   4096	\\
Halo mass               		& 150.0		&   40.0	&   7.0		\\
Halo cutoff radius          		& 200.0		&   80.0	&   45.0	\\
Halo core radius            		& 17.0		&   25		&   4.5		\\

   \tableline
   \end{tabular}
  \end{small}
 \end{center}
{\bf Note:} Simulations were done in a system of units with
G=1. Model units scale to physical ones such that: length unit is
1\,pc,  velocity unit is 65.58\,km\,s$^{-1}$,  mass unit is 
$1 \times 10^{6} \mathrm{M}_\odot$ and  time unit is
14909.92\,yr.

\end{table*}

Considering the range of uncertainties in the orbit determination,
we can state that this massive core would finally settle as the
new nucleus of M\,83 in a few tens of Myr, implying a net growth of
the central galactic mass. Furthermore, the whole star formation
and nuclei merging event would last about the time of a global
galactic revolution (about 150 Myr at the radius of 5 kpc).

Simulations also show that tidal-striping of the condensations boosts the
velocity field of the external shell to escape velocity, which, if mistaken for
the dispersion velocity of systems in equilibrium, might lead to an
overestimation of their masses. This seems to be the case for the disagreement
between the photometric and kinematic masses derived by
\citet{2000A&A...364L..47T} for ON.

The circumnuclear HII regions arc is far from being a stable system. In fact, it
will spread out in an orbital time and be swallowed by the new refurbished
nucleus in a period slightly longer than the time of merging of ON, KC and HN.
Therefore, the claimed existence of two inner Linblad resonances is difficult
sustain.

\section{Conclusion}\label{sec:conclu}

We performed CIRPASS 2D spectroscopy at Pa$\beta$ and its continuum with the
GEMINI-S telescope. Pa$\beta$ spectroscopy allows for penetration of the dust in
the direction of KC and HN, and for determination of the ionized gas kinematics
in the central $\approx 5\arcsec\times13\arcsec$ embracing ON, KC and HN. The
kinematics can mainly be explained in terms of three disks around each one of the
condensations, KC, HN and ON. Perturbations are visible when the three-disk
model is subtracted from the observations. The line of the nodes of a tens of 
parsecs-scale disk around KC is rotated with respect to that of the hundreds of 
parsecs-scale CO one, which in turn is rotated with regard to the line of the nodes of
the outer galactic disk. They are most probably coplanar up to the margin of error. 
Nevertheless, a relative warp of the galactic disk inwards of up to $40^{\circ}$
can not be ruled out.

Besides the mass of the condensations KC, HN and ON, we inferred upper mass
limits for the putative BH that could be associated to them.

Our simulation shows that the three nuclei, KC, HN and ON, will merge in a few
tens of Myr. The disruption of the circumnuclear HII regions arc in an
orbital time shows that even if ILR do exist, this is not enough to ensure the
stability of the HII regions' orbits.

We are witnessing a dramatic change in the central region of M83, the nearest
galaxy with strong nuclear star formation, which may in a few Myrs change the
mass at its kinematical center by a factor of 2, as well as the global aspect at
a hundred parsec scale. Although HN was interpreted as a captured dwarf
satellite, the metamorphosis of the M\,83 central region is independent of the
origin of HN.

\acknowledgements
I.R. and H.D. acknowledge support from CNPq (Brazil), I.R. also acknowledge
support from FAPESP (Brazil). R.D., M.A., and D.M. acknowledge support of
CONICET grant PIP 5697. We acknowledge support from Secyt-Capes Agreement,
project 035/07. We acknowledge the Instituto Nacional de Pesquisas Espaciais
(INPE/MCT, Brazil) for providing computer time for part of the simulations. Some
simulations were also run at IF-UFRGS, Brazil. The Gemini Observatory is
operated by the Association of Universities for Research in Astronomy, Inc.,
under a cooperative agreement with the NSF on behalf of the Gemini partnership:
NSF (USA), STFC (United Kingdom), NRC (Canada), ARC (Australia), MINCYT
(Argentina), CNPq (Brazil) and CONICYT (Chile). This publication makes use of
data products from the 2-MASS (Two Micron All Sky Survey), which is a joint
project of the University of Massachusetts and the Infrared Processing and
Analysis Center/California Institute of Technology, funded by the NASA and the
NSF. 

%\bibliography{references}

%\bibliographystyle{plainnat}

\end{document}